\documentclass[12pt]{article}
\let\a=\alpha\let\b=\beta\let\d=\delta
\let\e=\epsilon\let\g=\gamma
\let\th=\theta\let\l=\lambda
\let\m=\mu
\let\u=\upsilon

\let\S=\Sigma

\let\vt=\vartheta \let \L=\Lambda
\newcommand{\nn}{\nonumber}
\newcommand{\un}{\underline}
\newcommand{\uM}{\underline M}

\newcommand{\be}{\begin{equation}}
\newcommand{\ee}{\end{equation}}
\newcommand{\bea}{\begin{eqnarray}}
\newcommand{\eea}{\end{eqnarray}}
\newcommand{\eps}{\epsilon}

\renewcommand{\paragraph}[1]{
\vspace{.8mm}\par\noindent {\sl #1}\\
\vspace{0.2mm} }
\def\SLK{\hbox{\ooalign{$\displaystyle \Lambda_K$\cr$\hspace{2pt}/$}}}

\newcommand{\rep}{{\cal G}}
\newcommand{\eqn}[1]{(\ref{#1})}
\newcommand{\ft}[2]{{\textstyle\frac{#1}{#2}}}

\renewcommand{\u}[1]{{\bar{#1}}}
\newcommand{\ba}{\left(\begin{array}}
\newcommand{\ea}{\end{array}\right)}

\newcommand{\M}{{\cal M}}

\def\gg{{\cal G}}
\def\lie{{\cal L}}

\def\su{SU(2,2|4)}

\def\gM{\g^m}
\def\hg{\hat{\g}}

\def\pmatrixij{(\delta^{IJ} {\unity} + \hg'^{IJ})_i{}^j}
\def\mmatrixji{(\delta^{IJ} {\unity} - \hg'^{IJ})_j{}^i}

\def\bare{\bar{\e}}
\def\bareta{\bar{\eta}}
\def\bt{\bar{\theta}}

\def\bvt{\bar{\vt}}

\def\X{\Xi}
\def\adscft{$AdS$/$CFT$ }
\def\Miv{C_{(4|16)}}

\def\Mxxxii{C_{(10|32)}}

\def\adsxs{AdS_5 \times S^5}
\def\se{superembedding }

\newsavebox{\uuunit}
\sbox{\uuunit}
{\setlength{\unitlength}{0.825em}
 \begin{picture}(0.6,0.7)
\thinlines
\put(0,0){\line(1,0){0.5}}
\put(0.15,0){\line(0,1){0.7}}
 \put(0.35,0){\line(0,1){0.8}}
\multiput(0.3,0.8)(-0.04,-0.02){12}{\rule{0.5pt}{0.5pt}}
\end {picture}}
\newcommand {\unity}{\mathord{\!\usebox{\uuunit}}}

\begin{document}
\begin{titlepage}
\begin{flushright}
\vspace*{-1cm}
CALT-68-2284\\
CIT-USC/00-037\\
SU-ITP-00/18\\
{\tt hep-th/0007099}\\
\end{flushright}
\vskip 1cm

\begin{center}
{\large {\bf  Isometries in anti-de~Sitter
and Conformal Superspaces }}
\vskip 1
cm

{\bf  Piet Claus$^{1,a}$, J.
Rahmfeld$^{2,b}$, Harlan Robins$^{2,3,c}$, Jonathan
Tannenhauser$^{2,3,d}$, and Yonatan Zunger$^{4,e}$}\\
\vskip.3cm
{\small
$^1$Instituut voor theoretische fysica \\
Katholieke Universiteit Leuven, B-3001 Leuven, Belgium\\
\ \\

$^2$California Institute of Technology, Pasadena, CA 91125, USA \\

and\\

Caltech-USC Center for Theoretical Physics\\
University of Southern California, Los Angeles, CA 90089, USA\\
\ \\

$^3$Department of Physics \\ University of California at
Berkeley, Berkeley, CA 94720, USA\\
\ \\

$^4$Physics Department \\
Stanford University, Stanford, CA 94305-4060, USA\\
}\ \\

\vskip 0.2cm
\begin{abstract}
We derive explicit forms for the superisometries of a wide class
of supercoset manifolds, including those with fermionic
generators in the stability group. We apply the results to construct
the action of $\su$ on three supercoset manifolds:  $(10|32)$-dimensional
$AdS_5\times S^5$ superspace, $(4|16)$-dimensional conformal
superspace, and a novel $(10|16)$-dimensional conformal superspace.
Using superembedding techniques, we show, to lowest non-trivial order 
in the fermions, that at the boundary of
$AdS_5$, the superisometries of the $AdS_5 \times S^5$ superspace
reduce to the standard ${\cal N}=4$ superconformal transformations.  
In particular, half of the 32 fermionic coordinates
decouple from the superisometries.

\end{abstract}

\end{center}

\vfill

\footnoterule
\noindent
{\footnotesize
$\phantom{a}^a$piet.claus@fys.kuleuven.ac.be,
$\phantom{b}^b$rahmfeld@theory.caltech.edu,
$\phantom{c}^c$hrobins@theory.caltech.edu, \\
$\phantom{d}^d$jetannen@theory.caltech.edu,
$\phantom{e}^e$zunger@leland.stanford.edu }
\end{titlepage}

\section{Introduction \label{s:intro}}

A crucial element in the $AdS/CFT$ correspondence \cite{Malda}
is that the symmetry group $SU(2,2|4)$ is shared by string theory in 
$AdS_5\times S^5$ and ${\cal N} = 4$ super-Yang-Mills theory in four
dimensions.  In the string theory context, $\su$ appears as the group of
isometries of $C_{(10|32)}$, the $AdS_5 \times S^5$ superspace in
which the (Green-Schwarz) string propagates.  In the dual conformal field 
theory, $\su$ is the group of superconformal symmetries.  Equivalently, we may
take the gauge theory to be defined on a $(4|16)$-dimensional
conformal superspace, denoted $C_{(4|16)}$, whose superisometries---the
superconformal transformations---again form the group $\su$.  Both  
$C_{(10|32)}$ and $C_{(4|16)}$ are supercoset manifolds
of $\su$, and the realization of $\su$ on the coordinates of these spaces is
well studied.  

The $AdS/CFT$ correspondence identifies the boundary of anti-de~Sitter
space with the flat space on which the field theory is defined.  If we
are to implement the correspondence on the superspaces $C_{(10|32)}$
and $C_{(4|16)}$, then the isometries of $C_{(10|32)}$ must reduce, at the
boundary of $AdS_5$, to the superconformal
transformations of $C_{(4|16)}$.   In particular, six bosonic and 16
fermionic coordinates must decouple from the $C_{(10|32)}$
superisometries in the boundary limit.  
The decoupling of the six bosons is familiar
from the bosonic truncation of the 
$AdS/CFT$ correspondence.  What is new is the
decoupling of the fermions.  The purpose of this paper is to show,
to leading non-trivial order in an expansion in the fermionic
coordinates, that
the 32 fermions of the $AdS_5 \times S^5$ superspace $C_{(10|32)}$
split into sets of 16: one set may be
identified, at the boundary of $AdS_5$, with the fermionic coordinates
of $C_{(4|16)}$, and the other set decouples from the boundary superisometries.

Our first step is to derive explicit forms for the superisometries of
these superspaces.  In section 2, we develop general machinery for computing
supercoset isometries, extending the results of \cite{BlueBook,CK}.  Then
in section 3, we apply our general formulae to the supercosets
$C_{(10|32)}$ and $C_{(4|16)}$, and also to a novel conformal
superspace, denoted $C_{(10|16)}$, which is an enhancement of
$C_{(4|16)}$ by six extra bosonic coordinates. 
%
With our choice of supercoset representatives, the boundary
decoupling of the 16 fermionic coordinates
from the $C_{(10|32)}$ superisometries is not obvious.  We exhibit the
decoupling in section 4, by means of the superembedding approach of
\cite{Sorokin,Sezgin}.  
The superembedding approach offers an intuitive picture of the
boundary as a $(4|16)$-dimensional brane embedded in $C_{(10|32)}$.  
The geometry of the embedding is captured in a certain matrix (the {\em
\se matrix}), and the dynamics of the brane subjects this matrix to a 
constraint (the {\em \se equation}), 
whose solution provides us with the desired decomposition of bulk
fermions.  We find that the $C_{(10|32)}$ superisometries indeed reduce
at the boundary to the $C_{(4|16)}$ superconformal transformations,
with six bosonic and 16 fermionic coordinates decoupling. 


The identification of bulk and boundary coordinates has many potential 
applications. One example is the study of Wilson loops.  The $AdS/CFT$ 
correspondence relates the expectation value $\langle W({\cal C}) \rangle$ of
a Wilson loop operator along a contour ${\cal C}$ in the boundary 
${\cal N} = 4$ field theory to the action of the minimal-surface bulk
string worldsheet ending on the loop ${\cal C}$. This aspect of the 
correspondence has been investigated
mainly for purely bosonic boundary conditions \cite{Wilson1, reyyee, Wilson2,
Wilson3, Wilson4}. Extending the story to the fully supersymmetric
situation requires a detailed understanding of the relationship
between bulk and boundary coordinates in superspace. 
The present work is a step in this direction, and a full discussion will
be presented in \cite{SuperWilson}.  The $C_{(10|16)}$ superspace 
plays an important role: the additional six bosonic coordinates are
essentially identified
with the loop variables of \cite{Wilson4}, which appear in the definition of
the Wilson loop.

Another possible application is to the relation between gauged supergravity
and conformal supergravity \cite{TseytlinLiu, Nishimura, Transmo}.
For various maximal gauged supergravity theories with $AdS$ vacua,
it has been shown that the field transformation rules reduce on the 
boundary of the $AdS$ space to the transformation rules of conformal
supergravity in one lower
dimension \cite{Nishimura1, Nishimura,Nishimura2}. 
At the level of superspaces, this should be implemented
as an identification of the local superspace translations. 
The present work identifies the rigid limit of these superspace
translations in the case of five-dimensional
gauged ${\cal N}=8$ supergravity (in the $AdS_5$ vacuum) 
and four-dimensional ${\cal N}=4$ conformal supergravity.


\section{Superisometries of Coset Manifolds \label{s:genisom}}
\setcounter{equation}{0}

Our goal is to compare the isometries of $\adsxs$ superspace
and conformal superspace, both of which are supercoset manifolds of
the supergroup $SU(2,2|4)$.  In this section, we review a general
scheme for describing supercoset manifolds and their
isometries \cite{BlueBook,CK}, and obtain explicit expressions for the
isometries of a class of
coset supermanifolds that includes both  superspaces. 

Let $C = G / H$ be a coset manifold, with coordinates $Z^M =
(X^{\m}, \th^{\dot{\a}})$, and
let $\gg(Z) \in C$ be a coset representative.   Given $g \in G$, we can
always find an element $h(g,Z)$ of the stability
group $H$ such that 
\be
\label{cosetisom1}
\gg(Z) \rightarrow g \gg(Z) h^{-1}
\ee
is an isometry of $C$.  The meaning of this equation is as follows.
The element $g \in G$ defines an isometry of the group $G$ by
left-multiplication. 
However, in general this transformation is not an isometry of $C$:
if $\gg \in C$, $g \gg$ need not be an element of $C$.  In fact,  
\be
\label{compensating}
g \gg(Z) = \gg(Z') h,
\ee 
for some $Z'$ and some $h \in H$, with $h$  depending in general on
$g$  and $Z$.   To get back to $C$, we must compensate the
transformation  $g$ by right-multiplication by
$h$.  

Our characterization of the supercoset isometries will involve the
Cartan 1-forms 
\be
L(Z) = \gg(Z)^{-1} d \gg(Z), 
\ee
valued in the Lie algebra of $G$. The Cartan forms trivially satisfy
the Maurer-Cartan equation
\be 
dL - L \wedge L = 0.
\ee
They are left-invariant in the sense that they are unchanged under 
left-multiplication of $\gg(Z)$ by an element $g \in G$, 
\be
(g \gg(Z))^{-1} d (g \gg(Z)) - \gg(Z)^{-1} d \gg(Z) = 0.
\ee
Using \eqn {compensating}, we may rewrite this as
\be
\label{leftinvar}
h^{-1} L(Z') h + h^{-1} dh - L(Z) = 0.
\ee

Now let $g$ be an infinitesimal isometry.  Let the associated
coordinate transformation on the coset manifold\footnote{For
semi-simple groups $G$ and $H$, an isometry is equivalently
characterized by the invariance of the line element 
\be
g_{{M}{N}} dZ^M dZ^N= G_{{A}{B}}E^{A}_{M} E^{B}_{N} dZ^M dZ^N,
\label{eq:metrdef}
\ee
where $G_{{A}{B}}$ is the restriction to $C$ of the Cartan-Killing metric on
$G$, and $E$ is the vielbein of the coset manifold. However,
if $G$ or $H$ is not semi-simple---as is the case
for Minkowski superspace and for the conformal superspaces we shall
encounter in section 3---the 
Cartan-Killing metric is degenerate, and (\ref{eq:metrdef}) is not
well defined.  In this case, $g_{{M}{N}}$ must be replaced by an appropriate
$G$-invariant symmetric bilinear form, along the lines of
\cite{NappiWitten}.} be 
\be
-\X^M = Z'^M - Z^M \, ,
\ee
and let the infinitesimal compensating transformation be
\be
h = 1 + \L.
\ee
Then the statement \eqn {leftinvar} of the left-invariance of $L$ becomes
\be
\label{leftinvar2}
\lie_\X L + d \L + [L(Z), \L] = 0,
\ee
where $\lie_\X$ is the Lie derivative in the direction $\X$.
 
This is a $G$-covariant statement, but the form of \eqn {leftinvar2} is not
$G$-covariant: $L$ can take values in the full Lie superalgebra of $G$, but
$\L$ is restricted to lie in the Lie superalgebra of $H$.  To write \eqn 
{leftinvar2} in a $G$-covariant form, we define the quantity $\S =
\S^{\bar A} T_{\bar A}$ by 
\be
\label{Sdef}
\S^{A'} = \L^{A'} + \X^M L_M^{A'}, \qquad \S^{A} = \X^M L_M^{A}.
\ee
The notation is as in table~\ref{tab:algconv}: ${\bar A}$ runs over
all the generators of $G$, $A'$ indexes the generators
$H_{A'}$ of the stability group $H$, $A$ indexes the coset generators
$C_A$ ({\em i.e.}, the generators of $G$ which are not generators of $H$),
and $L = L^{\bar A} T_{\bar A}= L^{\bar A}_M dZ^M T_{\bar A}$.   
Using the Maurer-Cartan equations, we can rephrase \eqn
{leftinvar2} in the $G$-covariant form
\be
d \S + [L, \S] = 0\, ,
\ee
which is just the Killing equation for $\S$.
\begin{table}
\begin{center}
\begin{tabular}{|c|c|c|c|}
\hline
space&generators&bosonic generators&fermionic generators\\
\hline\hline
$G$ & $ T_{\bar A}$ & $T_{\bar a}$ & $T_{\bar \a}$\\
$C$ & $ C_A$ & $C_{a}$ & $C_{\a}$ \\
$H$ & $ H_{A'}$ & $H_{a'}$ & $H_{\a'}$\\
\hline
\end{tabular}
\end{center}
\caption{Conventions for the generators of $G$, $C$, and $H$.
\label{tab:algconv}}
\end{table}

We wish to solve for $\S$, $\X$, and $L$.  The
general solution of (2.12) is
\be
\S(Z) = \gg(Z)^{-1} \Upsilon \gg(Z),
\ee
where $\Upsilon$ is a constant element of the Lie superalgebra of
$G$.  To make further
progress, we must choose a parametrization of the coset
representative $\gg(Z)$.  We make the ansatz
\begin{equation}
\rep(Z) =  v(X) e^{\Theta}\,,
\end{equation}
where $v(X)$ is purely bosonic, $\Theta \equiv \Theta^\alpha
C_\alpha$, and  $\Theta^\alpha$
is related to the superspace fermionic coordinates $\theta^{\dot\alpha}$ by
\begin{equation}
\Theta^\alpha = \theta^{\dot \beta} t_{\dot \beta}{}^\alpha(X)\, .
\label{Thinth}
\end{equation}

We have the freedom to choose $t_{\dot \beta}{}^\alpha(X)$
arbitrarily.  This freedom amounts to a gauge choice in
superspace.  For now, we will leave $t_{\dot \beta}{}^\alpha(X)$
unspecified, and state our results for the isometries and the Cartan
form in full generality, but as we shall see in section 3, our formulae
simplify drastically with an astute choice of superspace gauge. 

The Cartan form and $G$-covariant Killing spinor are derived as in  
\cite{NearHorizon,CK},\footnote{Our problem is more general than the one in
\cite{NearHorizon, CK}, since  we allow $H$ to contain
fermionic generators. However, the methods used there to derive the 
Cartan forms and isometries are still valid.} with the result
\begin{eqnarray}
L^{\u \alpha} &=& \left((D\Theta) AB \right)^{\u\a},\qquad L^{\u a} =
L_0^{\u a} -  \left((D\Theta)A  \right)^{\u\b} \Theta^\alpha
f_{\alpha\u\beta}^{\u a} \,,\label{cartan2} \\
\Sigma^{\u \alpha} &=& \left(({\cal B}\Theta)AB\right)^{\u\alpha} +
\left( \Sigma_0 \M \coth \M
AB\right)^{\u\a}\,,\nonumber\\
\Sigma^{\u a} &=& \Sigma_0^{\u a} - \left(\Sigma_0 AB\right)^{\u\b} \Theta^\a
f_{\a\u\b}^{\u a}
- \left( ({\cal B} \Theta) A\right)^{\u\b} \Theta^\a f_{\a\u\b}^{\u a}\,,
\label{Gfields}
\end{eqnarray}
where
\begin{eqnarray}
A_{\u\a}{}^{\u\b} &=& 2\left( \frac{ \sinh^2 {\cal M}/2}{{\cal
M}^2}\right)_{\u\a}^{\ \ \ \u\beta}\,,\qquad B_{\u\a}{}^{\u\b}
= ({\cal M} \coth {\cal M}/2)_{\u\a}{}^{\u\b}\,,\nonumber\\
({\cal B}\Theta)^{\u\a} &=& \Theta^\beta \Sigma_0^{\u a} f_{{\u a}\beta}^{\u\a}\,,
\nonumber\\
\M^2_{\u\alpha}{}^{\u\beta} &=& f_{\u\alpha\gamma}^{\u a}
\Theta^\gamma\Theta^\delta
f_{\delta {\u a}}^{\u\beta}\, , \label{Msquare}
\end{eqnarray}
and the $f_{\u A \u B}^{\u C}$ are structure constants of the isometry
superalgebra.  The bosonic components of $L$ and $\Sigma$ involve
$L_0$ and $\Sigma_0$, the $\Theta = 0$ values of the
Cartan form and the Killing vector, 
\begin{equation}
L_0 = v(X)^{-1} d v(X) = e^{a} C_{a} + \omega^{a'} H_{a'}\,,\qquad
\Sigma_0 = v^{-1}(X) \Upsilon v(X)\ .
\end{equation}
We have introduced the quantity 
\be (D\Theta)^{\u \a} T_{\u \a}= d\Theta^\a C_\a +
dX^\mu  e_\mu{}^a U_{a}{}^{\u \b} T_{\u \b}\,, \label{DTheta}
\ee
where the matrix $U_{a}{}^{\u\alpha}$ depends on the superspace
gauge choice $t_{\dot\alpha}{}^{\beta}(X)$, and is given by
\be
U_{a}^{\u \a} T_{\u \a} = e_{a}{}^{\mu}\left(\partial_\mu \Theta
+ [L_0{}_\mu, \Theta] \right) = e_{a}{}^{\mu}\left(\theta^{\dot\alpha}
\partial_\mu
t_{\dot\alpha}{}^\alpha C_\alpha + \theta^{\dot\alpha}
t_{\dot\alpha}{}^\alpha L_{0\mu}^{\u a} f_{{\u a}\alpha}^{\u\b}
T_{\u\b}\right) \,. \label{Upart}
\ee

The last step in our program is the calculation of the superisometries
$\Xi$.  The Cartan form $L$  may be written as a linear combination of 
stability group
generators and coset generators,
\be
L = L^{\u A} T_{\u A} = E^A C_A + \Omega^{A'} H_{A'}.
\label{cartan}
\ee
The 1-form coefficients $E^A = E^A_M dZ^M$ are the vielbeins of the 
coset manifold, while the $\Omega^{A'}$ make up the $H$-{\it
connection}, 
which is the analogue for coset manifolds of the usual spin
connection. Inverting \eqn {Sdef}, we obtain
\be
\label{xiform}
\Xi^M=\Sigma^A (E^{-1})_A{}^M \, . 
\ee
We calculated $\Sigma^A$ in \eqn {Gfields}, but for \eqn {xiform} to be
truly useful, we need an expression for the inverse vielbein
coefficients.  We can compute the
vielbein itself relatively straightforwardly, by comparing
coefficients of the coset generators in \eqn{cartan} and
\eqn{cartan2}.  The result is
\begin{equation}
E_M{}^{A} = \ba{cc} e_\mu{}^b(X) & 0\\
0 & t_{\dot \alpha}{}^{\beta}(X) \ea \ba{cc} \delta_b{}^a +
(UA{\cal Y})_b{}^a & (UAB)_b{}^\alpha\\
(A{\cal Y})_\beta{}^a & (AB)_\beta{}^\alpha\ea\,, \label{svielbein}
\end{equation}
with
\begin{equation}
{\cal Y}_{\u\alpha}{}^a = - \Theta^\beta f_{\beta\u\a}{}^a\, .
\end{equation}

If the fermionic generators all lie in $C$, then  
\begin{equation}
(E^{-1})_A^M = \ba{cc} \delta_{a}{}^b & - U_{a}{}^{\beta}\\
-(B^{-1} {\cal Y})_{\alpha}{}^b & (B^{-1} {\cal
Y}U)_{\alpha}{}^{\beta} +
(AB)^{-1}{}_{\alpha}{}^{\beta}\ea \ba{cc} e_b{}^\mu & 0\\
0&t_{\beta}{}^{\dot\alpha}\ea\,
\label{eq:invviel1}
\end{equation}
and the superisometries are given by
\begin{eqnarray}
\Xi^\mu &=& \xi^\mu + \Sigma_0^{\a} (\M^{-1} \tanh \M/2 {\cal
Y})_{\a}{}^a
e_{a}{}^\mu\,,\nonumber\\
\Xi^{\dot \beta} t_{\dot\beta}{}^\alpha &=&
\left(\Theta^\beta \Sigma_0^{\u a} f_{\u a\beta}^\alpha - \xi^a
U_{a}{}^\alpha\right)\nonumber\\
&& + \Sigma_0^\beta (\M \coth \M)_\beta{}^\alpha -
\Sigma_0^\gamma (\M^{-1} \tanh \M/2)_\gamma{}^\beta
({\cal Y}U)_\beta{}^\alpha\,. \label{isogeneral}
\end{eqnarray}
All of the coset superspaces that arise as near-horizon limits of the
standard superbranes ($M2$, $M5$, $D3$, $D1$-$D5$), including $\adsxs$
superspace, are of this type. 

Inverting $E$ is more difficult if
the stability group contains fermionic generators. In this case, the
matrices $A$ and $B$ appearing in (\ref{svielbein}) are not square, so their
inverses, which are needed in (\ref{eq:invviel1}), are not well
defined. However, if the (anti)commutator of every fermionic coset generator
with every coset generator is itself in the coset, 
\be
\label{ccc}
[C_\a,C_A \}\in C,
\ee 
then we may still use (\ref{isogeneral}) to calculate the
superisometries, if we set the coordinates conjugate to the 
fermionic stability group generators to zero by hand.
Both of the
conformal superspaces we will study in section 3 contain fermionic
stability group generators, but satisfy the condition (\ref{ccc}).

\section{Superisometries of $\adsxs$ and Conformal Superspaces}
\setcounter{equation}{0}

We now apply the formalism of the last section to compute the 
superisometries of three distinct cosets of $G =
SU(2,2|4)$.\footnote{Our conventions for spinors and for the $SU(2,2|4)$ 
algebra are given in appendix A.}  The generators of $SU(2,2|4)$,
together with their weights under dilatations, are listed in table
\ref{tab:generators}. 
\begin{table}[hbtp]
\begin{center}
\begin{tabular}{|c|c|l|} \hline\hline
Operator&Weight&Name\\ \hline
$P_m$&1& Conformal Translations\\ \hline
$Q$&$1/2$& Global Supersymmetries\\ \hline
$M_{mn}$&0& Lorentz Rotations\\ \hline
$D$&$0$&Dilatation\\ \hline
$U^{~i}_j=(M'_{m'n'}, P'_{m'})$
&$0$&$SO(6)$ Rotations of $S^5$\\ \hline
$S$&$ - 1/2$&Special Supersymmetries\\ \hline
$K_{m}$&$- 1$&Special Conformal Transformations\\ \hline
\end{tabular}
\end{center}
\caption{$SU(2,2|4)$ generators in the superconformal basis.
The $SO(6)$ rotation $U$ is a linear combination of  
the $SO(5)$ rotation $M'_{m'n'}$  and the translation
$P'_{m'}$ ($m',n'=1, \dots, 5$) on the 5-sphere.}
\label{tab:generators}
\end{table}

As noted in the introduction, the three coset spaces we study are:

\noindent
(1) the $\adsxs$ superspace
\be
C_{(10|32)} = \frac{SU(2,2|4)}{SO(1,4) \times SO(5)},
\ee
in which the Green-Schwarz string in an anti-de~Sitter background
naturally propagates;

\noindent
(2) the conformal superspace
\be
C_{(4|16)} = \frac{SU(2,2|4)}{{\rm Span}(iso(1,3)_K \oplus D \oplus
so(6) \oplus S)},
\ee
on which the dual ${\cal N} = 4$ super-Yang-Mills theory may be
formulated; and

\noindent
(3) a novel conformal superspace
\be
C_{(10|16)} = \frac{SU(2,2|4)}{{\rm Span} (iso(1,3)_K \oplus so(5)
\oplus S)},
\ee
which differs from $C_{(4|16)}$ by the addition of six bosonic
coordinates. This superspace will find application in the study
of supersymmetric Wilson loops \cite{SuperWilson}.

The division of the $SU(2,2|4)$ generators into coset and
stability group generators for each coset space is shown in table
\ref{tab:cosets}.
\begin{table}[hbtp]
\begin{center}
\begin{tabular}{|c||c|c|c|} \hline
&$C_{(10|32)}$&$C_{(10|16)}$&$C_{(4|16)}$\\ \hline\hline
&&&\\
$C$&$\frac{1}{2}(P_m+K_m),P'_{m'},D, Q, S$& $P_m,~~
P'_{m'}, ~~ D, ~~ Q$ & $P_m, ~~~~~~Q$ \\ 
 & & & \\ \hline
&&&\\
$H$& $\frac{1}{2}(P_m-K_m), M_{mn},
M'_{m'n'} $ & $K_{m}, M_{mn},
M'_{m'n'}, S$& $K_{m}, D, P'_{m'}, M_{mn},
M'_{m'n'}, S$  \\
&&& \\\hline
\end{tabular}
\end{center}
\caption{Coset decompositions of $SU(2,2|4)$ Generators
\label{tab:cosets}}
\end{table}

\subsection{Superisometries of $\adsxs$ superspace \label{ss:adSisom}}

{\sl Super-horospheric coordinates} \\ 
We begin by describing our choice of coordinates on the $\adsxs$
superspace $C_{(10|32)}$.  For bosonic coordinates on $AdS_5$ we
choose the {\em horospheric} coordinates $(x^m, \rho)$, which appear
naturally in the near-horizon limit of brane solutions in
supergravity. The $x^m$ are four Cartesian bulk 
coordinates parallel to the brane, while $\rho$ is the bulk coordinate
perpendicular to the brane. In these coordinates the
$AdS_5$ metric has the form
\begin{equation}
ds^2 = \rho^2 dx^2 + \left(\frac R\rho\right)^2 d\rho^2\,,
\end{equation}
where $R$ is the characteristic length (``radius'') of $AdS_5$. In the following we
will set $R=1$.  The boundary of $AdS_5$ is located
at $\rho = \infty$.  The coordinates on $S^5$ are denoted by
$\phi^{m'}$. Sometimes we group
the coordinates of the sphere and the radial coordinate of $AdS_5$ 
into a system of $6$ Cartesian coordinates $y^I$, with
\begin{equation}
\rho = |y|\,.
\end{equation}

We wish to extend these bosonic coordinates to a coordinate system on
$C_{(10|32)}$ in which the superisometries take as simple a form as
possible.  This involves fixing the arbitrary matrix $t_{\dot
\beta}{}^{\alpha}$ in \eqn{Thinth}.  To guide our choice, we notice
that all fermionic dependence in (\ref{cartan2})-(\ref{Upart}) is
polynomial in $D\Theta$ and $\Theta$.  The most desirable choice of 
$t_{\dot\beta}{}^{\alpha}(X)$ would simultaneously simplify $D\Theta$
and $\Theta$, and so render the calculation of the superisometries tractable.

The most obvious choice $t_{\dot\beta}{}^{\alpha}(X) =
\d_{\dot\beta}{}^{\alpha}$ (Wess-Zumino gauge) simplifies $\Theta$,
but $D\Theta$ remains complicated.  Another option is to choose 
$t_{\dot\beta}{}^{\alpha}$ 
such that, if $\eps^{\dot \a}$ is a constant spinor, the spinor
\be
\eps^{\a}(X)=\eps^{\dot \beta}t_{\dot\beta}{}^{\a}(X),
\ee
satisfies the Killing equation
\be
(\d_\b^\a d+L^{\bar a}f_{\bar a \b}^\a)\eps^{\b}(X)=0\, .
\ee
This ensures that
\be
\label{Killgauge}
U_{a}{}^{\bar\alpha} = 0.
\ee
In this gauge, called Killing gauge \cite{NearHorizon}, 
\be
(D \Theta)^\a = (d\theta)^{\dot \b} t_{\dot \b}{}^\a \qquad
\hbox{and} \qquad
(D \Theta)^{\a'}=0 \, . 
\ee
However, with this choice the expressions for the superisometries
are not translationally invariant, and $\Theta$ itself is a rather
messy function of $\theta$. We therefore work in a different 
gauge, which we now introduce.

Let us write the fermionic generators of $SU(2,2|4)$ in the superconformal
decomposition (for details see appendix~A.2)
\begin{equation}
 {\cal Q}_{\alpha}{}^i = \pmatrix{ Q_\alpha{}^i\cr
S_{\alpha}{}^i}.
\end{equation}
For the fermionic parametrization of the coset we choose
\begin{eqnarray}
\Theta &=& \bar \Theta_i {\cal Q}^i + \bar {\cal Q}_i \Theta^i \nonumber\\
&=& (u^{-1})_i{}^j \bar \theta_i \rho^{1/2} Q^i
+ (u^{-1})_i{}^j \bar \vt_j\rho^{-1/2} S^i
+ \bar Q_i \rho^{1/2} \theta^j u_j{}^i
+ \bar S_i \rho^{-1/2} \vt^j u_j{}^i\,.
\label{QSdecomp}
\end{eqnarray}
Here $u_i{}^j= u_i{}^j(\phi)$ is a coset representative of
$SO(6)/SO(5)$; the indices $i,j$ are spinor indices of $SO(6)$.  
We name the gauge defined by (\ref{QSdecomp}) {\em
partial-Killing gauge}, as it is intermediate between the Wess-Zumino and
Killing gauges.  In this gauge,
\be
U_{\rho}^{\bar \a}=U_{m'}^{\bar \a}=0\, ,
\ee
but $U_{\m}^{\bar \a}$ need not equal zero; the Killing equation is
satisfied in the $S^5$ and radial $AdS$
directions, but not in the directions parallel to the boundary.
The coordinates  
$Z^M = \{x^m, \rho, \phi^{m'}, \theta^i, \vt^i,
\bar \theta_i, \bar \vt_i \}$ are called {\em super-horospheric} coordinates.

\bigskip
\noindent
{\sl The $AdS_5\times S^5$ supergeometry and isometries}\\
The supercoset $C_{(10|32)}$ is maximally supersymmetric; the
stability group contains no fermionic generators.  The isometries are
calculated from \eqn{isogeneral}.

The geometry is given by the supervielbein
\begin{equation}
E = E^{\u M} C_{\u M} =
E^m P_m + E^\rho D + E^{m'} P_{m'} + (\bar Q_i
E_Q^i + \mbox{h.c.}) + (\bar S_i E_S^i + \mbox{h.c.})\,,
\end{equation}
where
\begin{eqnarray}
E^m &=& \rho \left[dx^n\left(\delta_n{}^m - \ft 12\left(\ft
1\rho\right)^2 \bar\vt_i \gamma_n \vt^j\bar\vt_j \gamma^m
\vt^i\right) +  \left(\ft12 d \bar\theta_i \gamma^m \theta^i +
\ft1{4} \bar\theta_i d\vt^j \bar\theta_j \gamma^m \theta^i +
\mbox{h.c.}\right) \right. \nonumber\\ &&\phantom{\rho
\left[\right.} + \left(\ft 1\rho\right)^2 \left.\left( \ft12
d\bar\vt_i \gamma^m \vt^i + \ft1{4} \bar\vt_i d\theta^j \bar\vt_j
\gamma^m \vt^i + \mbox{h.c.}\right)\right] + {\cal
O}(\theta\land\vt)\,, \nonumber\\ E^\rho &=& \ft 1\rho\left[ d\rho
- \ft12\left(d\bar\theta_i \vt^i - d\bar\vt_i \theta^i +
\mbox{h.c.}\right)\rho \right] + {\cal O}(\theta\land\vt)\,,
\nonumber\\ E^{m'} &=& e^{m'} - i \ft 12\left(d\bar\theta_i \vt^j
+ d\bar\vt_i \theta^j + dx^m \bar\vt_i \gamma_m \vt^j -
\mbox{h.c.}\right) \left( u \gamma'{}^{m'} u^{-1}\right)_j{}^i +
{\cal O}(\theta\land\vt) \,,\nonumber\\ E_Q^i
&=&\rho^{1/2}\left[d\theta^j - dx^m \gamma_m \vt^j + \ft13
\theta^k\left(d\bar\vt_k\theta^j - \bar \theta_kd\vt^j\right)
\right] u_j{}^i + {\cal O}(\theta\land\vt)\,,\nonumber\\ E_S^i
&=&\rho^{-1/2}\left[d\vt^i + \ft13\vt^k\left(2d\bar\theta_k\vt^j -
\bar\vt_k d\theta^j\right) + dx^m \vt^k\bar\vt_k\gamma_m \vt^j
\right] u_j{}^i + {\cal O}(\theta\land\vt)\,. \label{vielbeinadS}
\end{eqnarray}
Here ${\cal O}(\theta\land\vt)$ stands for terms containing both
$\theta^i$ and $\vt^i$.  
As we have noted, it is sometimes convenient
to replace the coordinates $(\rho, \phi^{m'})$ by Cartesian
coordinates $y^I$.  In these coordinates, $E^\rho$ and
$E^{m'}$ are subsumed into
\begin{eqnarray}
E^I &=& \frac1\rho\left[dy^I - \ft12 \left((d\bar\theta_i \vt^j +
dx^m \bar\vt_i \gamma_m \vt^j )(\delta^{IJ}\unity  - \hat
\gamma'^{IJ})_j{}^i y_J + \mbox{h.c.}\right)\right. \nonumber\\
&&\left.\phantom{\frac1\rho\left[ dy^I\right. } + \ft12
\left(d\bar\vt_i \theta^j (\delta^{IJ}\unity  + \hat
\gamma'^{IJ})_j{}^i y_J + \mbox{h.c.}\right)\right] + {\cal
O}(\theta\land\vt)\,.
\end{eqnarray}

The superisometries then follow from substituting into (\ref{isogeneral}):
\begin{eqnarray}
\delta x^m &=& -\xi^m_C(x)
               - \ft12 \left(\bar\epsilon_i(x) \gamma^m \theta^i +
\mbox{h.c.}\right) - \ft1{4} \left(\bar\eta_i\theta^j
\bar\theta_j\gamma^m \theta^i + \mbox{h.c.}\right)\nonumber\\ &&-
\left(\ft R\rho\right)^2\left[\Lambda_K^m + \ft12 \left(\bar\eta_i
\gamma^m \vt^i + \mbox{h.c.}\right) + \ft1{4}
\left(\bar\epsilon(x)_i\vt^j \bar\vt_j\gamma^m \vt^i +
\mbox{h.c.}\right)\right]\nonumber\\ && + {\cal
O}(\theta\land\vt)\,,\nonumber\\
\delta \rho &=& \Lambda_D(x) \rho + \ft12\left(\bar \epsilon_i(x)  \vt^i -
\bar \eta_i\theta^i +\mbox{h.c.}\right) \rho + {\cal
O}(\theta\land\vt)\,,\nonumber\\
\delta \phi^{m'} &=& - \xi^{m'}(\phi) + \ft i2\left(\bar \epsilon_i(x)  \vt^j +
\bar \eta_i\theta^j - \mbox{h.c.}\right)\left(u
\gamma'{}^{m'}u^{-1}\right)e_{\u m'}{}^{m'}
+ {\cal O}(\theta\land\vt)\,,\nonumber\\
\delta y^I &=& \Lambda_D(x) y^I - \Lambda_{SO(6)}^{IJ}\, y_J + \ft12\left(
\bar\epsilon_i(x) \vt^j (\delta^{IJ}\unity  - \hat \gamma'^{IJ})_j{}^i y_J +
\mbox{h.c.}\right)\nonumber\\
&& - \ft12 \left( \bar\eta_i \theta^j (\delta^{IJ}\unity +
\hat\gamma'{}^{IJ})_j{}^i y_J + \mbox{h.c.}\right)
 + {\cal O}(\theta\land\vt)\,,\nonumber\\
\delta \theta^i &=& - \epsilon^i(x) - \ft12 \Lambda_D(x) \theta^i - \ft14
\Lambda_M(x) \cdot \gamma \theta^i - \ft14 \theta^i
\Lambda^{IJ}_{SO(6)}(\hat\gamma'_{IJ})_j{}^i\nonumber\\
&& - \left(\ft R\rho\right)^2 \left[ \Lambda_K^m + \ft12\left(\bar \eta_j
\gamma^m \vt^j + \mbox{h.c.}\right) + \ft1{4} \left(\bar\epsilon_j(x) \vt^k
\bar\vt_k \gamma^m \vt^j + \mbox{h.c.}\right) \right] \gamma_m \vt^i
\nonumber\\
&& - \ft23 \theta^j(2\bar\eta_j \theta^i - \bar\theta_j \eta^i)
+ {\cal O}(\theta\land\vt)\,,\nonumber\\
\delta\vt^i &=& - \eta^i + \SLK \theta^i + \ft12 \Lambda_D(x) \vt^i -
\ft14 \Lambda_M(x) \cdot \gamma \vt^i - \ft14 \vt^i
\Lambda^{IJ}_{SO(6)}(\hat\gamma'_{IJ})_j{}^i\nonumber\\
&&- \ft23 \vt^j\left(2\epsilon_j(x) \vt^j  - \bar\vt_j\epsilon^i(x)\right)
 + {\cal O}(\theta\land\vt)\,.
\label{adSSisom}
\end{eqnarray}

We have written these transformations in terms of the $x$-dependent
parameters of superconformal transformations,
\begin{eqnarray}
\xi_C^m(x) &=& a^m + \lambda_M^{mn} x_n + \lambda_D x^m + (x^2
\Lambda_K^m - 2 x^m x\cdot \Lambda_K)\,,\nonumber\\
\Lambda_M^{mn} (x) &=& \lambda^{mn}_M - 4 x^{[m}
\Lambda_K^{n]}\,,\nonumber\\
\Lambda_D (x) &=& \lambda_D - 2 x\cdot \Lambda_K\,,\nonumber\\
\Lambda_K^m &=& \Lambda_K^m\,,\nonumber\\
\epsilon^i(x) &=& (\eps^i + x^m \gamma_m \eta^i)\,.
\label{confisom}
\end{eqnarray}
Here $a^m, \lambda_M^{mn},\lambda_D, \Lambda_K^m$ are
the constant parameters of translations, Lorentz rotations,
dilatations and special conformal transformations; $\eps$
and $\eta$ parametrize supersymmetries and special supersymmetries; and
the $\Lambda_{SO(6)}^{IJ}$ are the parameters of the $SO(6)$ $R$-symmetry.


In the boundary limit $\rho \to \infty$, the radial coordinate $\rho$ decouples
from the bosonic isometries $\d x^{\m}$, and these isometries reduce
at the boundary to the conformal transformations of conformal
space. We likewise expect that 16 fermions decouple from the
superisometries in the boundary limit, and that the superisometries
reduce in that limit to the superconformal transformations of
conformal superspace.  The decoupling of the fermions is not evident
in (\ref{adSSisom}), though.  In fact, from (\ref{adSSisom}) it would seem
that the boundary isometries 
depend on all 32 fermionic coordinates.  We will resolve this
puzzle in section 4, by exhibiting a change of coordinates in which
the decoupling of 16 fermions is apparent.  To make the decoupling precise, we
need the form of the superconformal transformations of the boundary.
We now study these transformations in the context of the novel
conformal superspace $C_{(10|16)}$ defined above.

\subsection{The $C_{(10|16)}$ superisometries \label{ss:newCS}}

The novel conformal superspace $C_{(10|16)}$ is an enlargement of  
the conformal Minkowski superspace $C_{(4|16)}$ by the six bosonic
coordinates $v^I = (v, \varphi^{m'})$.  At the level of the $SU(2,2|4)$
algebra, this enlargement is realized by reclassifying the generator $D$ 
(corresponding to $\rho$) and the generators of $SO(6)/SO(5)$
(corresponding to the sphere coordinates) as coset generators, rather
than as stability group generators.  Fermionically, $C_{(10|16)}$
differs from that of $C_{(10|32)}$ in that
generators $S$ are assigned to the stability group.  The 16 fermionic 
coordinates are defined by
\begin{equation}
\Theta = (u^{-1})_i{}^j \bar \l_i v^{1/2} Q^i
+ \bar Q_i v^{1/2} \l^j u_j{}^i\,,
\end{equation}
which is the natural adaptation of (\ref{QSdecomp}).  Within this
coset this choice corresponds to the Killing gauge.
  
The coordinates $z^M$ of $C_{(10|16)}$ conjugate to the generators 
$(P_m,
Q^i, D, P_{m'}')$ are $(w^m, \l_i^{\a}, v, \varphi^{m'})$.
The supervielbein $e$  
\begin{equation}
e = e^m P_m - e^v D + e^{m'} P_{m'} + (\bar Q_i e^i + \mbox{h.c.})\,
\end{equation}
of $C_{(10|16)}$ may be obtained either by direct calculation or by
setting $\vt = d \vt = 0$, $(x^m, y^I, \th^i) = (w^m, v^I, \l^i)$ in the Cartan form of
$AdS_5\times S^5$ superspace. The result is essentially the
vielbein of flat superspace.
The isometries are given by
\begin{eqnarray}
\delta w^m  &=&  -\xi^m_C(w) -
\ft12(\bar\epsilon_i(w) \gamma^m \l^i + \hbox{h.c.}) + \ft12 (\bar\l_i\eta^j - \bar \eta_i\l^j)
\bar\l_j\gamma^m
\l^i - \ft12 \bar\l_i\SLK \l^j \bar\l_j \gamma^m
\l^i\,,\nonumber\\
\delta \l^i &=& -\epsilon^i (w) -\ft12 \Lambda_D(w)
\l^i -
\ft14 \Lambda_M\cdot \gamma \l^i -\ft14\l^j
\Lambda_{SO(6)}^{IJ} (\hat\gamma'_{IJ})_j{}^i  -  \l^j
(2\bar\eta_j\l^i - \bar \l_j \eta^i) \nn \\ && - 
\ft12 \l^j \bar \l_j \SLK \l^i \, ,\nonumber\\
\delta v^I &=& \Lambda_D(w) v^I - \Lambda_{SO(6)}^{IJ} v_J - \left((\bar \eta_i\l^j + \ft12 \bar\l_i \SLK
\l^j)
(\unity \delta^{IJ} + \hat \gamma'{}^{IJ})_j{}^i v_J + \mbox{h.c.} \right)\,.
\label{CSisom}
\end{eqnarray}
These transformations are precisely the superconformal
transformations derived in \cite{Ferber} via supertwistors,
extended to the bulk directions $v^I$.
They realize the $SU(2,2|4)$ algebra on the 10+16 coordinates,
as desired. 

The $v^I$ coordinates are completely decoupled from the transformations of $x$
and $\theta$.  Their associated generators can therefore be freely
moved into the stability group, to give $C_{(4|16)}$, without
affecting the isometries.  Indeed, $\d w^m$ and $\d \l^i$ in
(\ref{CSisom}) are exactly the superisometries of conformal Minkowski 
superspace.

\section{$AdS_5\times S^5$ Superisometries
and Superconformal Transformations}
\setcounter{equation}{0}

In the bosonic truncation
of the \adscft correspondence, the isometries of the bulk space
restrict at the boundary to the conformal symmetries of
the boundary theory.  Any sensible extension of \adscft to superspace
must preserve this feature; namely, the superisometries of the bulk space must
reduce in the boundary limit to the superconformal transformations of
conformal superspace.  In particular, in the example we studied in
section 3, 16 of the 32 fermionic coordinates of $\Mxxxii$ must
decouple from the superisometries in the limit $\rho \to \infty$.  But
no decoupling of this sort is apparent in \eqn {adSSisom}, nor is it clear what
the proper relationship is between the $\Miv$ fermionic coordinates
$\l$ and the $\Mxxxii$ fermions $\theta$ and $\vt$.

We can state the difficulty another way.  If the \adscft
correspondence extends to superspace, then it must be
possible to carve out of the $10+32$ coordinates of $AdS_5 \times S^5$
superspace a set 
of $4+16$ supercoordinates $z^M=(w^m,\lambda^i)$, with 
\bea
w^{m} &=& w^{ m}(x^{m}, y^I, \th, \vt), \nn \\
\l &=& \l(x^{m}, y^I, \th, \vt), 
\label{coordtransf}
\eea
and with the property that
\be
\label{isomatch}
\d_{(10|32)} z^M|_{\rho = \infty} = \d_{(4|16)} z^M.
\ee
Here $\d_{(10|32)}$ and $\d_{(4|16)}$ denote the superisometries \eqn
{adSSisom} and the superconformal transformations \eqn {CSisom}.  We
are thus asking for a $(4|16)$-dimensional brane $C_{(4|16)}$
embedded in $\Mxxxii$ at $\rho = \infty$, on which the
superisometries of $\Mxxxii$ restrict to the superisometries of conformal 
superspace. The equations \eqn {isomatch}
are overconstrained, so the existence of a solution is non-trivial.  
 
Nonetheless, we will find one.  Our method uses the superembedding
formalism of \cite{Sorokin,Sezgin}.  Consider a superbrane with
coordinates $z^M$ and vielbein $e_M{}^{\un M}$, embedded in a
larger target superspace with coordinates $Z^{\bf M}$ and vielbein $E_{\bf
M}{}^{\un {\bf M}}$.\footnote{Underlined and non-underlined indices
refer to local Lorentz and general coordinate frames, respectively.}
The geometry of the embedding is encoded in the
{\em \se matrix} 
\be
\hat E_{\uM}{}^{\un{\bf M}}  \equiv (e^{-1})_{\uM}{}^M \partial_M
Z^{{\bf M}} E_{{\bf M}}{}^{\un{\bf M}}\, .
\ee
The dynamics of the embedded superbrane are contained in the
requirement that the \se matrix satisfy the {\em \se equation} 
\be
\label{superembedding}
\hat{E}_{\un{{\a}}}{}^{\bf \un {\mu}} = 0 \,.
\ee
where $\m$ runs over the bosonic directions of the target space.   The
\se equation states that the fermionic tangent space of the superbrane lies 
entirely within the fermionic tangent space of the target space, and
does not protrude into the bosonic part of the tangent space of the target.

For us, the embedded superbrane is the boundary space $C_{(4|16)}$,
and the target space in which it is embedded is the bulk space
$C_{(10|32)}$.  The \se matrix is given in terms of the vielbein of
$C_{(10|32)}$,\footnote{Extracting non-trivial information from the \se
matrix requires more terms in the expansion of the vielbein of $C_{(10|32)}$ 
than were presented in section 3.  In Appendix B
we give the bulk vielbein and superisometries to quartic order in the
fermions, in the relevant limit $\rho \to \infty$.} the
unknown coordinate transformation \eqn
{coordtransf}, and the vielbein of $C_{(4|16)}$, which is just the
vielbein of flat superspace,
\bea
\label{boundvielbein}
e^{\un m} &=& dw^{m} \d_{m}^{\un m} + \ft 1 2 d \bar{\l}_i \g^{\un m} \l^i -
\ft 1 2 \bar{\l}_i \g^{\un m} \l^i \,, \nn \\
e^{\un{i}} &=& d \l^{\un i} \,.  
\eea
Thus the \se equation constrains the coordinate transformation \eqn
{coordtransf}.  The most naive identifications $w^m=x^m, \ \lambda^i=\theta^i$
violate (\ref{superembedding}). However, the corrected transformations
\bea
\label{redefs}
w^m&=& x^m - \ft 1 4 (\bar{\th}_i \vt^j - \bar{\vt}_i \th^j)
(\bar{\th}_j \g^{\m} \th^i) \, , \nn \\
\l_\a{}^i&=& \theta_{\a}{}^i + 
\ft{1}{3} \theta_{\a}{}^j (2 \bar\vartheta_j \theta^i - \bar\theta_j
\vartheta^i)  
\eea
solve \eqn {superembedding} to third order in $\th$ and $\vt$.
We expect that an expansion like \eqn {redefs} can be developed to all
orders $\th$ and $\vt$.\footnote{In \cite{SuperWilson}, a different
coset representative is used to prove (\ref{isomatch}) to all orders, 
and the corrections in (\ref{redefs}) are attributed to Baker-Hausdorff 
rearrangements.} 
Moreover, the coordinate transformations (\ref{redefs}) satisfy
(\ref{isomatch}): 16 fermionic degrees of freedom do decouple from the  
$C_{(10|32)}$ superisometries in the limit $\rho \to \infty$, and the
superisometries indeed reduce in that limit to
the superisometries of conformal superspace. 


The redefinitions (\ref{redefs}) can be extended to a map from
$C_{(10|32)}$ to the novel conformal superspace $C_{(10|16)}$, with 
coordinates $(w^m, v^I, \lambda^i)$, by supplementing (\ref{redefs})
with the transformations 
\be
\label{redefsE}
v^I = y^I - \frac{1}{2} 
\bar \vt_i \theta^j \mmatrixji y_J -  \frac{1}{2}\bar \theta_j \vt^i
\pmatrixij y_J \,. 
\ee
Again, the superisometries of the bulk restrict at the boundary to the
superconformal transformations of the novel conformal superspace, in
the sense of (\ref{isomatch}).

\bigskip

\bigskip

\noindent
{\bf Acknowledgements:} We have enjoyed useful discussions with
Djordje Minic,  Hirosi Ooguri and Dmitri Sorokin.
We are particularly thankful to Renata Kallosh for comments and extensive 
collaboration during many stages of this
work. J.R. was supported by the Caltech Discovery Fund and
DE-FG03-92-ER40701. H.R. and
J.T. are supported in part by NSF grant PHY-95-14797 
and DOE grant DE-AC03-76SF00098. Y.Z. was partially supported by an
NSF graduate research fellowship.

\newpage

\appendix

\section{The $SU(2,2|4)$ Algebra and Spinors in $AdS_5\times S^5$ and 
Conformal Superspace}
\setcounter{equation}{0}

The $SU(2,2|4)$ algebra contains $SO(2,4)$ as a bosonic subalgebra.
We begin by discussing the Dirac matrices of $SO(2,4)$ and $SO(6)$.  
We index the 
$SO(2,4)$ directions by $\hat m = \{ m,S,T \}$, with $m = 0,\dots,3$
and signature $\hat \eta_{\hat m \hat n} = {\rm diag}(-++++-)$.
Written in a chiral basis, the $SO(2,4)$ Dirac matrices have the form
\begin{equation}
\hat \Gamma_m = \ba{cc} 0&\gamma_m\\\gamma_m &0\ea\,,\quad \hat
\Gamma_S = \ba{cc} 0&\gamma_5\\\gamma_5 &0\ea\,,\quad \hat
\Gamma_T = \ba{cc} 0&-1\\1&0\ea\,,
\end{equation}
where the $\gamma_m$ are the Dirac matrices of $SO(1,3)$, and
$\gamma_5 = i \gamma_0 \gamma_1 \gamma_2 \gamma_3$.  

The chiral decomposition proves useful, since the minimal spinor in
$AdS_5 \times S^5$ dimensions is chiral with respect to both $SO(2,4)$
and $SO(6)$.  We
will work with right-handed spinors $\l$ satisfying $\hat \Gamma_7 \l
= - \l$, where 
\be
\Gamma_7 = i \hat \Gamma_0 \hat \Gamma_1 \cdots \hat \Gamma_S \hat
\Gamma_T = \ba{cc}1 
&0\\0&-1 \ea \,.
\ee
We denote the restriction of $\hat \Gamma$ to the right-handed chiral
spinor subspace by $\hat \g$.  The antisymmetrized products of the 
$\hat \g_{\hat m}$ are given by
\begin{equation}
\label{chiralgamma}
\hat \gamma_{mn} = \gamma_{mn}\,,\quad \hat\gamma_{mS} =
\gamma_m\gamma_5\,,\quad \hat \gamma_{mT} = -\gamma_m\,,\quad \hat
\gamma_{TS} = \gamma_5\,.
\end{equation}
These matrices satisfy the Fierz identities
\begin{equation}
(\hat\gamma_{\hat m\hat n})_{\alpha}{}^{\beta}
(\hat\gamma^{\hat m\hat n})_{\gamma}{}^{\delta} = 2
\delta_{\alpha}{}^{\beta}
\delta_{\gamma}{}^{\delta} - 8
\delta_{\alpha}{}^{\delta}\delta_{\gamma}{}^{\beta}\,, \qquad
(\hat \gamma_{\hat m\hat n})_{\alpha}{}^{\beta}
(\hat\gamma^{\hat p\hat q})_{\beta}{}^{\alpha} = - 8
\delta_{[\hat m}{}^{\hat p} \delta_{\hat n]}{}^{\hat q}\,.
\label{fierz}
\end{equation}

The $SO(6)$ directions are indexed by $I$.  We denote the Dirac
matrices of $SO(6)$, restricted to the chiral subspace, by $\hat \gamma'^{I}$.
If we divide the index $I$ as $I=\{ T',m' \}$, where $m' = 1,\dots,5$ runs 
over the $S^5$ directions, and consider antisymmetrized products as in 
(\ref{chiralgamma}), then the matrix $\gamma'_{m'}$ appearing in 
(\ref{vielbeinadS}) and (\ref{adSSisom}) is
given by $\gamma'_{m'} = \hat \gamma'_{m'T'}$.  

The group $SU(2,2|4)$ is generated by: $SO(2,4)$ transformations $\hat
M_{\hat m \hat n}$ ($\hat m, \hat n = 0,\dots,3,S,T$); $SU(4)$
transformations $U_i^j$ ($i,j = 1,\dots,4$); and fermionic
transformations ${\cal Q}_{\alpha}^i$ ($\alpha = 1,\dots,4$) and
${\cal Q}_i^{\alpha} = i \left(({\cal Q}^i)^{\dag} \gamma_0
\right)^{\alpha}$. The
structure of the algebra is
\begin{eqnarray}
{}[\hat M_{\hat m\hat n}, \hat M_{\hat p\hat q}] &=& \hat
\eta_{\hat m[\hat p} \hat M_{\hat q]\hat n} - \hat \eta_{\hat
n[\hat p} \hat M_{\hat q]\hat m}\,, \nonumber\\ {}[\hat M_{\hat
m\hat n}, {\cal Q}_{\a}{}^i] &=& -\frac 14 
(\hat \gamma_{\hat m\hat n} {\cal Q}^i)_{\alpha}\,,\nonumber\\ \{
{\cal Q}_{\alpha}{}^i, \bar {\cal Q}_j{}^{\beta} \} &=&
\frac12 \delta_j{}^i (\hat \gamma^{\hat m\hat
n})_{\alpha}{}^{\beta} \hat M_{\hat m\hat n} -
\delta_{\alpha}{}^{\beta} U_j{}^i\,,\nonumber\\
{}[U_i{}^j, {\cal Q}_{\alpha}^k] &=&  \delta_i{}^k
{\cal Q}_{\alpha}{}^j - \frac 1{4} \delta_i{}^j {\cal
Q}_{\alpha}{}^k\,, \nonumber\\ {}[U_i{}^j, U_k{}^l] &=&
\delta_i{}^l U_k{}^j -  \delta_k{}^j U_i{}^l\,,
\label{SU224inSO}
\end{eqnarray}
plus complex conjugates. All other commutators vanish. 

We can rewrite this in two ways, corresponding to two different
breakdowns of the $SO(2,4)$ subalgebra.

\subsection{The $AdS$ Decomposition}
The $AdS$ decomposition takes note of the coset structure
\be
AdS_5 = SO(2,4)/SO(1,4)\, .
\ee
We split the generators of $SO(2,4)$ into 
\be
\tilde P_{\tilde m}=2 \hat M_{{\tilde m}T}, \qquad \tilde M_{\tilde m\tilde n}= 
\hat M_{\tilde m \tilde n}\,, 
\ee
with $\tilde m, \tilde n=0,1,2,3,4=S$. The $\tilde M_{\tilde m\tilde
n}$ generate the stability $SO(1,4)$, while the $\tilde P_{\tilde m}$
generate translations in $AdS_5$. In this decomposition,
the $SU(2,2|4)$ algebra is
\begin{eqnarray}
{}[\tilde M_{\tilde m\tilde n}, \tilde M_{\tilde p\tilde q}] &=& \tilde
\eta_{m[\tilde p} \tilde M_{\tilde q]\tilde n} - \tilde \eta_{\tilde
n[\tilde p} \tilde M_{\tilde q]\tilde m}\,, \nonumber\\ 
{}[\tilde P_{\tilde q},\tilde M_{\tilde m \tilde n}]= \tilde \eta _{\tilde q [\tilde m}
\tilde P_{\tilde n]},&& {}[\tilde P_{\tilde m},\tilde P_{\tilde n}]=2 \tilde M_{\tilde m
\tilde n} \nn \\
{}[\tilde M_{\tilde
m\tilde n}, {\cal Q}_{\a}{}^i] &=& -\frac 14 
(\hat \gamma_{\tilde m\tilde n} {\cal Q}^i)_{\alpha}\,,\nonumber\\
{}[\tilde P_{\tilde
m}, {\cal Q}_{\a}{}^i] &=& \frac 12 
(\hat \gamma_{\tilde mT} {\cal Q}^i)_{\alpha}\,,\nonumber\\
 \{
{\cal Q}_{\alpha}{}^i, \bar {\cal Q}_j{}^{\beta} \} &=&
\frac12 \delta_j{}^i (\hat \gamma^{\tilde m T})_{\alpha}{}^{\beta} 
\tilde P_{\tilde m}
+\frac12 \delta_j{}^i (\hat \gamma^{\tilde m\tilde
n})_{\alpha}{}^{\beta} \tilde M_{\tilde m\tilde n} - 
\delta_{\alpha}{}^{\beta} U_j{}^i\,,\nonumber\\
{}[U_i{}^j, {\cal Q}_{\alpha}^k] &=&  \delta_i{}^k
{\cal Q}_{\alpha}{}^j - \frac 1{4} \delta_i{}^j {\cal
Q}_{\alpha}{}^k\,, \nonumber\\ {}[U_i{}^j, U_k{}^l] &=&
\delta_i{}^l U_k{}^j -  \delta_k{}^j U_i{}^l\,,
\label{SU224AdS}
\end{eqnarray}
plus complex conjugates.  All other commutators vanish.

\subsection{The Superconformal Decomposition}
In the superconformal decomposition of $SU(2,2|4)$, the
$SO(2,4)$ subalgebra is decomposed by weight under the dilatation operator 
\be
D=2 \hat M_{TS}\, .
\ee
The four-dimensional translations 
\be
P_m=2 (\hat M_{mT}+\hat M_{mS})
\ee
($m=0,1,2,3$) have weight $1$ ({\it i.e.}, $[D, P_m] = P_m$); the
$SO(1,3)$ Lorentz transformations
\be
M_{mn}=\hat M_{mn}\, ,
\ee
the $SU(4)$ transformations $U_i{}^j$ and the dilatation
operator $D$ itself have weight zero; and the special
conformal transformations
\be
K_m=2 (\hat M_{mT}-\hat M_{mS})
\ee
have weight $-1$.
The fermionic operators split into supersymmetries
$Q$, with weight $1/2$, and special supersymmetries $S$, with weight
$-1/2$, according to
\begin{equation}
Q_\alpha^i = 
\frac{1}{\sqrt{2}}(1-\gamma_5)_{\alpha}{}^{\beta}{\cal Q}_{\beta}{}^i\,,\quad
S_\alpha^i = \frac{1}{\sqrt{2}} 
(1+\gamma_{5})_{\alpha}{}^{\beta}{\cal Q}_{\beta}{}^i\,.
\end{equation}
That is, $D$ divides the fermionic generators according to their
$SO(1,3)$ chirality.
In this decomposition, the algebra becomes
\begin{eqnarray}
{}[ M_{ m n},  M_{ p q}] &=& 
\eta_{m[ p}  M_{ q] n} -  \eta_{
n[ p}  M_{ q] m}\,, \nonumber\\ 
{}[ P_{ q}, M_{ m  n}]=  \eta _{ q [ m}
 P_{ n]},&& {}[ K_{ q}, M_{m n}]=\eta _{ q [ m}
 K_{ n]}, \nn \\
{}[D, P_{m}]= P_{ m},&& {}[D, K_{m}]=-K_{m}, \nn \\
{}[ P_{m}, K_{n}]&=&2\left(\eta_{mn} D+2 M_{mn}\right),\nn \\
{}[ M_{
m n}, Q_{\a}{}^i] = -\frac 14 
(\gamma_{ m n} Q^i)_{\alpha}\,, & & 
{}[ M_{
m n}, S_{\a}{}^i] = -\frac 14 
(\gamma_{ m n} S^i)_{\alpha}\, ,
\nonumber\\
{}[ P_{
m}, S_{\a}{}^i] =  
(\gamma_{m} Q^i)_{\alpha}\,, &&
{}[ K_{
m}, Q_{\a}{}^i] =  
(\gamma_{m} S^i)_{\alpha}\, ,
\nonumber\\
{}[D, Q_{\a}{}^i] =\frac{1}{2}  
Q^i_{\alpha}\,, &&
{}[D, S_{\a}{}^i] =-\frac{1}{2}  
S^i_{\alpha}\,, 
\nonumber\\
 \{
{Q}_{\alpha}{}^i, \bar {Q}_j{}^{\beta} \}=\d_j{}^i (\gamma^m)_\a{}^\b P_m\, ,
&&
\{
{S}_{\alpha}{}^i, \bar {S}_j{}^{\beta} \}=\d_j{}^i (\gamma^m)_\a{}^\b K_m\, , \nn \\
\{
{Q}_{\alpha}{}^i, \bar {S}_j{}^{\beta} \} &=&
\delta_j{}^i \delta_{\alpha}{}^{\beta} D +\delta_j{}^i (\gamma^{ mn})_{\alpha}{}^{\beta}M_{mn} 
-2 
\delta_{\alpha}{}^{\beta} U_j{}^i\,,\nonumber\\
{}[U_i{}^j, {Q}_{\alpha}^k] &=&  \delta_i{}^k
{Q}_{\alpha}{}^j - \frac 1{4} \delta_i{}^j {Q}_{\alpha}{}^k\,, \nn \\
{}[U_i{}^j, {S}_{\alpha}^k] &=&  \delta_i{}^k
{S}_{\alpha}{}^j - \frac 1{4} 
\delta_i{}^j {S}_{\alpha}{}^k\,, \nonumber\\ {}[U_i{}^j, U_k{}^l] &=&
\delta_i{}^l U_k{}^j -  \delta_k{}^j U_i{}^l\, ,
\label{SU224SC}
\end{eqnarray}
plus complex conjugates. All other commutators vanish. 

\subsection{Fermionic Coordinates of  $SU(2,2|4)$ Cosets}
A general minimal spinor in $AdS_5\times S^5$ is
chiral with respect to $SO(2,4)$ and $SO(6)$, and so contains 32
degrees of freedom.  We write the 32 fermionic
coordinates of the $AdS_5\times S^5$ superspace $C_{(10|32)}$ as
$\Theta_\a^i,  \bar \Theta_i^{\alpha}$,
where  $\a=1,2,3,4$ and  $i=1,2,3,4$ are the $SO(2,4)$
and $SO(6)$ spinor indices.  As in the discussion above, the conjugate spinor 
${\bar \Theta}_i^\a$ is defined by
\be
{\bar \Theta}_i^\a=i\left((\Theta^i)^\dagger\gamma_0\right)^\a\, .
\ee
The superconformal decomposition splits the fermionic
generators with respect to their $SO(1,3)$ chirality, which brings
about a split of the coordinate $\Theta_\a^i$ into
\bea
\theta_\a^i&=& \frac{1}{2\sqrt{2}}\left((1+\gamma_5)\Theta^i\right)_\a\,, \nn \\
\vartheta_\a^i&=& \frac{1}{2\sqrt{2}}\left((1-\gamma_5)\Theta^i\right)_\a\, .
\eea
The $\theta_\a^i$ are the coordinates conjugate to the supersymmetries $Q$,
and the $\vartheta_a^i$ are conjugate to the special supersymmetries $S$.

In the conformal superspaces $C_{(4|16)}$ and $C_{(10|16)}$, the
$Q_\a^i$ are elements of the coset and have conjugate coordinates
$\th_\a^i$, but the $S_\a^i$ are elements of the stability group and
have no corresponding conjugate coordinates.

\section{The superisometries of $AdS_5\times S^5$ to quartic 
order in fermions}
\setcounter{equation}{0}

We present to quartic order in the fermions the bosonic vielbein $E^m$
of $C_{(10|32)}$ and the coordinate variations under supersymmetries
and  special supersymmetries, in the limit $\rho \to \infty$.  These
are needed to establish the coordinate transformations 
(\ref{redefs})-(\ref{redefsE}) and the corresponding
matching of superisometries and superconformal transformations.  We have
explicitly calculated the matching only for the supersymmetries and special
supersymmetries, but we expect the matching to hold for all superisometries.

\bea
E^m &=& \rho \left[ dx^n \delta_n{}^m + \ft 1 2  \left( (d \bar \th_i  + \bar \vt_i  \gamma_n dx^n) \gamma^m \theta^i 
+  \ft 1 4 (\bar \theta_i d \vt^j)(\bar \theta_j \gamma^m \theta^i)
\right. \right. \nn \\ && \left. \left. + \ft 1 4 \bar \vt_j(d
\theta^i - dx^n \gamma_n \vt^i) (\bar \theta_i \gamma^m \theta^j) +
\ft 1 {12} (\bar \theta_j \gamma^m (d \theta^i - dx^n \gamma_n
\vt^i))(\bar \theta_i \vt^j + \bar \vt_i \theta^j) \right. \right. \nn
\\ && \left. \left. + \hbox{h.c.} \right) \right],  \\ 
\delta x^m &=& -\ft{1}{4}(\bare_j \vt^i \bt_i \gM \theta^j + 
\bareta_j \th^i \bt_i \gM \th^j - \bvt_i \e^j \bt_j \gM \theta^i -
 \bt_i \eta^j \bt_j \gM \th^i) \nonumber \\ && - \ft{1}{12} 
(\bare_j \gM \theta^i \bt_i \vt^j + \bare_j \gM \theta^i \bvt_i \theta^j - 
\bt_i \gM \e^j \bvt_j \theta^i - \bt_i \gM \e^j \bt_j \vt^i), \\
\delta \theta^i 
&=& - \eps^i - \ft43 (\bare_j \vt^i  + \bareta_j \th^i) \theta^j +
 \ft23 \left( \e^j(\bt_j \vt^i + \bvt_j \theta^i) + (\bt_j \eta^i +
 \bvt_j \e^i) \th^j \right) - \nn \\ &&
- \ft12 (\bare_j \g^a \theta^j  - \bt_j \g^a \e^j)
 (\g_a \vt)^i, \\
\d y^I &=& \ft12 \left[ \bar \epsilon_i \vt^j (\d^{IJ} \unity - \hat
\gamma'^{IJ})_j{}^i + \bar \vt_i \epsilon^j (\d^{IJ} \unity +  
\hat \gamma'^{IJ})_j{}^i - \bar \eta_i \th^j (\d^{IJ} \unity + 
\hat \gamma'^{IJ})_j{}^i - \right. \nn \\ && \left. - 
\bar \th_i \eta^j (\d^{IJ} \unity - 
\hat \gamma'^{IJ})_j{}^i \right] y^J.
\eea

\end{document}